\documentclass{aastex631}

\shorttitle{C$^{18}$O and C$^{17}$O J=2-1 Data Analysis}
\shortauthors{Zou et al.}

\usepackage{CJK}%Chinese
\usepackage{hyperref}
\hypersetup{hypertex=true,
            colorlinks=true,
            linkcolor=blue,
            anchorcolor=blue,
            citecolor=blue}
\usepackage{longtable}
\usepackage{multirow}
\usepackage{threeparttable}
\usepackage{pgffor}
\usepackage{subfigure}
\usepackage{amsmath}

\defcitealias{Zhang2020}{Paper~I}

\begin{document}
\begin{CJK*}{UTF8}{gbsn}%Chinese
\title{A Systematic Observational Study on Galactic Interstellar Ratio $^{18}$O/$^{17}$O. II. C$^{18}$O and C$^{17}$O J=2-1 Data Analysis}
\author[0000-0002-5230-8010]{Y. P. Zou (邹益鹏)}
\affiliation{Center for Astrophysics, Guangzhou University, 
Guangzhou, 510006, People's Republic of China}

\author[0000-0002-5161-8180]{J. S. Zhang (张江水)}
\affiliation{Center for Astrophysics, Guangzhou University, 
Guangzhou, 510006, People's Republic of China}
\correspondingauthor{J. S. Zhang}
\email{jszhang@gzhu.edu.cn}

\author[0000-0002-7495-4005]{C. Henkel}
\affiliation{Max-Planck-Institut f{\"u}r Radioastronomie, Auf dem H{\"u}gel 69, D-53121 Bonn, Germany}
\affiliation{Astronomy Department, King Abdulaziz University, P.O. Box 80203, 21589 Jeddah, Saudi Arabia}
\affiliation{Xinjiang Astronomical Observatory, Chinese Academy of Sciences, 830011 Urumqi, Peopleʼs Republic of China}

\author[0000-0002-0845-6171]{D. Romano}
\affiliation{INAF, Osservatorio di Astrofisica e Scienza dello Spazio, Via Gobetti 93/3, I-40129 Bologna, Italy}

\author{W. Liu (刘玮)}
\affiliation{Center for Astrophysics, Guangzhou University, 
Guangzhou, 510006, People's Republic of China}

\author[0009-0009-2303-395X]{Y. H. Zheng (郑映慧)}
\affiliation{National Astronomical Observatories, Chinese Academy of Sciences, Beijing, China}
\affiliation{University of Chinese Academy of Sciences, Beijing, China}

\author{Y. T. Yan (闫耀庭)}
\affiliation{Max-Planck-Institut f{\"u}r Radioastronomie, Auf dem H{\"u}gel 69, D-53121 Bonn, Germany}

\author[0000-0001-8980-9663]{J. L. Chen (陈家梁)}
\affiliation{Center for Astrophysics, Guangzhou University, 
Guangzhou, 510006, People's Republic of China}

\author{Y. X. Wang (汪友鑫)}
\affiliation{Center for Astrophysics, Guangzhou University, 
Guangzhou, 510006, People's Republic of China}
\author{J. Y. Zhao (赵洁瑜)}
\affiliation{Center for Astrophysics, Guangzhou University, 
Guangzhou, 510006, People's Republic of China}
\begin{abstract}
To investigate the relative amount of ejecta from high-mass versus intermediate-mass stars and to trace the chemical evolution of the Galaxy, we have performed with the IRAM 30 m and the SMT 10 m telescopes a systematic study of Galactic interstellar $^{18}$O/$^{17}$O ratios toward a sample of 421 molecular clouds, covering a galactocentric distance range of $\sim$1 - 22 kpc. The results presented in this paper are based on the J=2-1 transition and encompass 364 sources showing both C$^{18}$O and C$^{17}$O detections. The previously suggested $^{18}$O/$^{17}$O gradient is confirmed. For the 41 sources detected with both facilities, good agreement is obtained. A correlation of $^{18}$O/$^{17}$O ratios with heliocentric distance is not found, indicating that beam dilution and linear beam sizes are not relevant. For the subsample of IRAM 30 m high-mass star-forming regions with accurate parallax distances, an unweighted fit gives $^{18}$O/$^{17}$O = (0.12 $\pm$ 0.02)$R_{GC}$ + (2.38 $\pm$ 0.13) with a correlation coefficient of $R=0.67$. While the slope is consistent with our J=1-0 measurement, ratios are systematically lower. This should be caused by larger optical depths of C$^{18}$O 2-1 lines, w.r.t the corresponding 1-0 transitions, which is supported by RADEX calculations and the fact that C$^{18}$O/C$^{17}$O is positively correlated with $^{13}$CO/C$^{18}$O. After considering optical depth effects with C$^{18}$O J=2-1 reaching typically an optical depth of $\sim$0.5, corrected $^{18}$O/$^{17}$O ratios from the J=1-0 and J=2-1 lines become consistent. A good numerical fit to the data is provided by the MWG-12 model\deleted{ from \cite{Romano2019}}, including both rotating stars and novae.
\end{abstract}
\keywords{Interstellar molecules (849); Radio sources (1357); Isotopic abundances (867); Galaxy chemical evolution (580)}

\section{Introduction} \label{sec:intro}
The evolution of metallicity in the Galactic disk is the consequence of stellar nucleosynthesis, which converts hydrogen to heavier elements, which are subsequently ejected into the interstellar medium (ISM; e.g., \citealt{Wilson1994}).
\replaced{Radial metallicity gradients in the Galactic disk have been found in different objects, including stars (e.g., \citealt{Xiang2017}), H II regions (e.g., \citealt{Esteban2018}), and planetary nebulae (e.g., \citealt{Henry2010}), supporting the inside-out formation scenario for our Galaxy \citep{Larson1976}.}
{{The presence of Galactic radial metallicity gradients in various objects, such as stars (e.g., \citealt{Xiang2017}), H II regions (e.g., \citealt{Esteban2018}), and planetary nebulae (e.g., \citealt{Henry2010}), supports the inside-out formation scenario of our Galaxy \citep{Larson1976}.}}
\replaced{As particularly suitable tracers of the nucleosynthesis and stellar ejecta are isotope abundance ratios}{{Isotope abundance ratios serve as particularly suitable tracers for nucleosynthesis and stellar ejecta}}, because they are not only addressing overall elemental abundances but are focusing instead on specific isotopes. These can be effectively measured by observations of molecular clouds in the radio, mm- and submm-bands, through the analysis of molecular species with more than one stable isotopologue (e.g., \citealt{Yan2019, Humire2020, Yu2020, Chen2021, Yan2023}).

Observing specific isotopes, the interstellar $^{18}$O/$^{17}$O ratio is one of the most useful tracers of nuclear processing and metal enrichment.
\replaced{$^{18}$O is believed to be primarily synthesized in massive stars ($M\geq8M_\odot$) by helium burning on $^{14}$N, while $^{17}$O should be predominantly ejected by intermediate mass stars through carbon-nitrogen-oxygen (CNO) burning on a longer timescale (e.g., \citealt{Henkel1993}). In this case (i.e., with a longer production timescale for $^{17}$O), the $^{18}$O/$^{17}$O ratio from the Galactic center (GC) region should be smaller than that from the Galactic disk, assuming an inside-out formation scenario for our Galaxy (\citealt{Matteucci1989}).}
{{It is widely accepted that $^{18}$O and $^{17}$O have different nucleosynthetic paths. Briefly, $^{18}$O is primarily synthesized in massive stars, while $^{17}$O is predominantly ejected by intermediate mass stars through a longer production timescale. This can lead to a mildly positive gradient of $^{18}$O/$^{17}$O along the disk, assuming an inside-out formation scenario for our Galaxy (see detailed description in \citealt{Zhang2020}).}}
\deleted{In fact, recent calculations show that the nucleosynthesis of $^{18}$O and $^{17}$O in stars is more complex than the picture sketched above. A significant production of primary $^{18}$O from fast-rotating massive stars may occur at low metallicity, while nova systems may contribute non-negligible amounts of $^{17}$O to their surroundings (see the discussion in \citealt{Romano2019}).}\replaced{These}{{Recent}} theoretical results mainly based on numerical calculations suggest a complex interdependence of the yields on mass, metallicity and rotation of the stars. The resulting large uncertainties require observational data to constrain such models. \deleted{$^{18}$O/$^{17}$O ratios are readily determined from C$^{18}$O/C$^{17}$O line intensity ratios, because C$^{18}$O and C$^{17}$O have similar chemical and excitation properties and tend to be optically thin. Furthermore, oxygen isotopes are not plagued by significant fractionation effects (e.g., \citealt{Langer1984}), which is related to the high first ionization potential of oxygen. Finally, intensity ratios of the C$^{18}$O and C$^{17}$O lines are typically well below a factor of ten, thus requiring observational sensitivities that are not extremely different.}

\replaced{Previous measurements on the isotopic ratio of $^{18}$O/$^{17}$O were mainly performed toward individual sources, through observing the C$^{18}$O and C$^{17}$O lines (e.g., \citealt{Bensch2001, Ladd2004, Wouterloot2005, Zhang2007}). \cite{Penzias1981} firstly investigated variations of the isotopic ratio toward a sample of 15 molecular clouds and reported one uniform value for $^{18}$O/$^{17}$O, varying less than $\sim$5\% between the GC and disk molecular clouds with galactocentric distance out to 12 kpc. However, \cite{Wouterloot2008} suggested a radial gradient toward a sample of 18 sources along the galactic disc, from the GC to galactocentric distance out to 16 kpc.
More data from the GC region and far outer parts of the Galaxy were suggested to improve the statistical significance of their results. Furthermore, mainly thanks to the Bar and Spiral Structure Legacy (BeSSeL) Survey\footnote{http://bessel.vlbi-astrometry.org} \citep{Reid2014, Reid2019}, distances and thus locations inside the Galaxy are much more accurately defined than 15 years ago.}
{{$^{18}$O/$^{17}$O ratios can be readily determined from C$^{18}$O/C$^{17}$O line intensity ratios (see details, e.g., \citealt{Zhang2007}). Previous measurements on the isotopic ratio were mainly performed toward individual sources (e.g., \citealt{Bensch2001, Ladd2004, Wouterloot2005, Zhang2007}) or small samples \citep{Penzias1981, Wouterloot2008}. Measured results of $^{18}$O/$^{17}$O seems to be contradictive, one uniform value between the Galactic center (GC) and disk molecular clouds \citep{Penzias1981} or a radial gradient along the galactic disc \citep{Wouterloot2008}. A large sample is really important to determine the ratio and its possible trend, especially more data from the GC region and far outer parts of the Galaxy.}}

Therefore, we started a systematic observational study based on the $^{18}$O/$^{17}$O isotope ratio, measuring rotational transitions of C$^{18}$O and C$^{17}$O. \replaced{Our earlier work includes C$^{18}$O and C$^{17}$O mapping toward molecular clouds in the GC for the first time \citep{Zhang2015} and a single-pointing pilot survey of Galactic disk molecular clouds (14 sources) with different distances \citep{Li2016}, which showed lower abundance ratios toward the GC region with respect to molecular clouds in the Galactic disk, though the ratios appear not to be uniform inside the GC region.}{{Our earlier work indicated lower abundance ratios in the GC region compared to molecular clouds in the Galactic disk, based on mapping toward GC molecular clouds and a single-pointing pilot survey of Galactic disk molecular clouds \citep{Zhang2015, Li2016}.}}
The first article in our series of $^{18}$O/$^{17}$O studies involving large samples of sources was based on measurements of the J = 1-0 lines of C$^{18}$O and C$^{17}$O (\citealt{Zhang2020}, hereafter \citetalias{Zhang2020}). It confirmed the previously suggested Galactic radial gradient of $^{18}$O/$^{17}$O. However, one transition alone is not sufficient to account for radiative transfer effects. Reliable statistical results on the abundance ratio $^{18}$O/$^{17}$O still need more transition lines to evaluate potential optical depth effects and thus to obtain more reliable isotope ratios.
Here we present the so far missing J = 2-1 data from C$^{18}$O and C$^{17}$O. As mentioned in \citetalias{Zhang2020}, our large sample includes star formation regions (SFRs) associated with IRAS sources and relatively strong CO emission ($T_A^* \gtrsim$ 10 K, \replaced{selected from \citealt{Wouterloot1989}}{{with kinematic distances taken from \citealt{Wouterloot1989}}}) and high-mass star-forming regions (HMSFRs), whose distances have been accurately measured by the maser \replaced{parallax method \citep{Reid2014}. In addition, more HMSFRs with trigonometric parallax distance were added, thanks to recently reported results from the BeSSeL Survey \citep{Reid2019}.}{{parallax method from the Bar and Spiral Structure Legacy (BeSSeL\footnote{http://bessel.vlbi-astrometry.org/}) Survey \citep{Reid2014, Reid2019}.}} Our sample includes 421 sources, covering galactocentric distances from the GC region to the far outer Galaxy ($\sim$22 kpc). \added{{Among them, 200 sources have accurate trigonometric parallax distance with a median relative uncertainty of less than 10\% (according to their measured parallax data).}}
The observations with the IRAM 30 m and the SMT 10 m are described in Section \ref{sec:obs}, the data reduction and corresponding results are presented in Section \ref{sec:d&r}, while Section \ref{sec:dsc} contains an analysis and the discussion. Main results are summarized in Section \ref{sec:sum}.

\section{Observations}\label{sec:obs}
\subsection{IRAM 30 m observations}
Our observations of the C$^{18}$O and C$^{17}$O J=2-1 lines were carried out from 2017, Jan. 11 till Jan. 17, using the IRAM 30 m, at the Pico Veleta Observatory (Granada, Spain). The center frequencies were set at 219.560354 and 224.714187 GHz for the C$^{18}$O and C$^{17}$O lines, respectively. The observations were performed in position switching mode with the off position $30\arcmin$ from the source. Using the Eight Mixer Receiver (EMIR) with dual-polarization and the Fourier Transform Spectrometers (FTS) backend, a frequency coverage of 218-226 GHz in the lower sideband was obtained, with a spectral resolution of 195 kHz (corresponding to a velocity resolution of 0.26 km s$^{-1}$) around 222 GHz. The system temperature was about 250 K with an rms noise \replaced{of about 44 mK}{{range in the resulting spectra of 21-119 mK with an median value of 38 mK}} on the antenna temperature scale $T_{A}^{*}$. Antenna temperatures can be transformed, multiplied by the ratio of the forward hemisphere efficiency and the main beam efficiency ($F_{eff}/B_{eff} \sim 0.92/0.59 = 1.56$\footnote{https://publicwiki.iram.es/Iram30mEfficiencies}) into main beam brightness temperatures, $T\rm_{mb}$.

With the IRAM 30 m, we observed 103 sources of our sample. Observational parameters are summarized in Table \ref{tab:results}. The source name and its equatorial J2000 coordinates are listed in Columns 1-3. The galactocentric and heliocentric distance of each source, the telescope used, and targeted molecular species are listed in Columns 4-7. Column 8 provides the rms noise of observations, Column 9 is \replaced{the velocity of each component}{{the velocity of the peak $T\rm_{mb}$}}, and Columns 10 and 11 list the integrated line intensity and the peak temperature, respectively, obtained from Gaussian fits to the spectra. For each source the results for C$^{18}$O are on the first and those of C$^{17}$O are on the second line. Column 12 presents the abundance ratios (see details in Section \ref{sec:d&r}).
\subsection{SMT 10 m observations}
Our SMT 10 m (on Mt. Graham, AZ, USA) observations of J=2-1 lines were performed remotely in 2016 May and June, 2018 January and November, 2019 December, 2020 June, July and November, as well as 2021 January and February with a beam size of $\sim$29\arcsec. A dual-polarization 1.3 mm receiver frontend and the SMT 10 m filter bank backends in the 2 IF mode were used, which provide bandwidths of 1000 and 256 MHz and spectral resolutions of 1000 and 250 kHz (corresponding to velocity resolutions of 1.32 and 0.33 km s$^{-1}$, respectively. The latter was used for our analysis, in order to make comparisons with IRAM 30 m data). A position switching mode with reference positions $30\arcmin$ off in azimuth was adopted. On the antenna temperature scale $T_{A}^{*}$, the system temperature was $\sim$380 K with an rms noise \replaced{of $\sim$55 mK}{{range in the resulting spectra of 14-533 mK with an median value of 37 mK}} in our observations. The antenna temperature can be converted, divided by the main beam efficiency correction factor ($\eta_{b} = 0.71$\footnote{https://aro.as.arizona.edu/?q=beam-efficiencies}), into main beam brightness temperature $T\rm_{mb}$.

Table \ref{tab:results} presents the observational parameters of the SMT 10 m results for the sample of 380 sources, including those 62 observed by both the IRAM 30 m and the SMT 10 m telescopes. Comparing observational results of the same source from different telescopes is necessary to evaluate calibration uncertainties and the effect of the beam size on the line ratios.

\section{Data reduction and Results}\label{sec:d&r} 
\begin{figure}[ht]
\centering
\includegraphics[width=\textwidth,page=1]{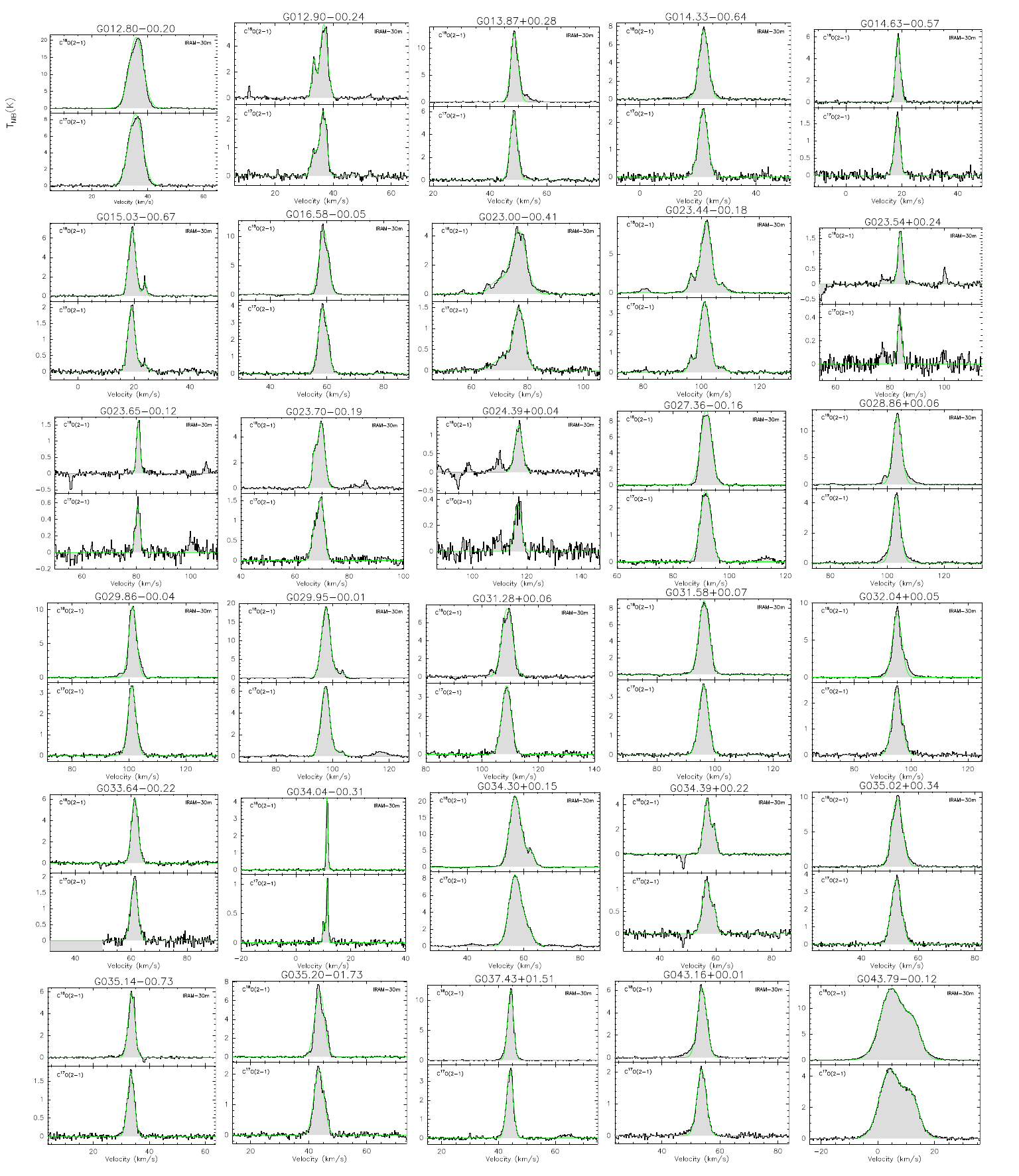}
\caption{The IRAM 30 m spectra of C$^{18}$O (upper panels) and C$^{17}$O (lower panels) with green fit lines of the 96 sources detected in both isotopologues.\\
(An extended version of this figure is available).}
\label{fig:iram_both}
\end{figure}

\begin{figure}[ht]
\centering
\includegraphics[width=\textwidth,page=1]{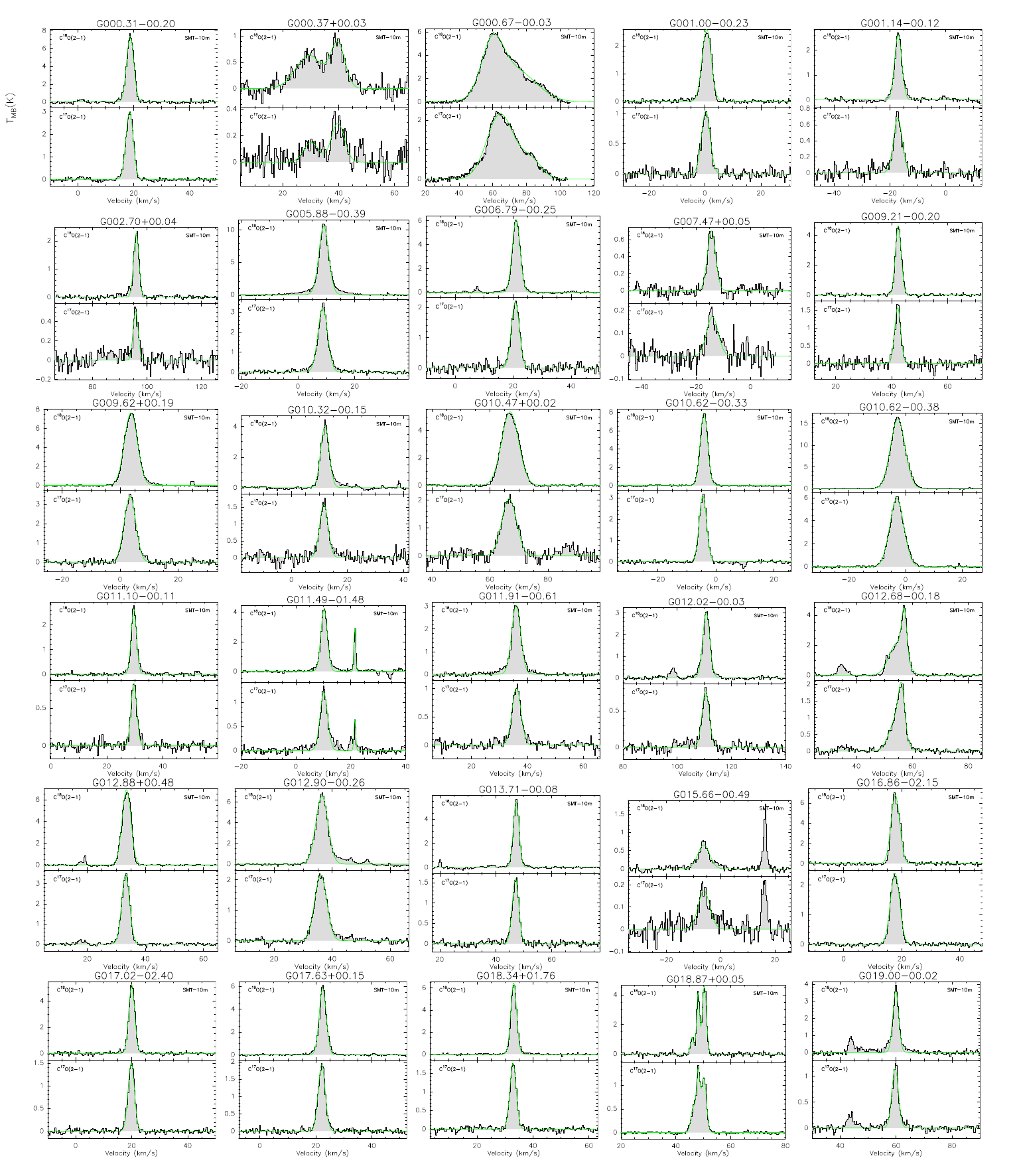}
\caption{The SMT 10 m spectra of C$^{18}$O (upper panels) and C$^{17}$O (lower panels) with green fit lines of the 325 sources detected in both isotopologues.\\
(An extended version of this figure is available).}
\label{fig:smt_both}
\end{figure}

Data reduction was done using the Continuum and Line Analysis Single-dish Software (CLASS) of the Grenoble Image and Line Data Analysis Software packages (GILDAS). 
Baselines were subtracted with polynomial fitting. After baseline subtraction, line parameters were obtained from Gaussian fits to the C$^{18}$O and C$^{17}$O spectra (see green lines in Figures \ref{fig:iram_both} and \ref{fig:smt_both}).

96 out of 103 sources were successfully detected in both C$^{18}$O and C$^{17}$O J=2-1 by the IRAM 30 m, while 325 out of 380 sources were detected in both lines using the SMT 10 m. Moreover, 57 out of 62 common sources observed by both telescopes were detected in both lines. To summarize, 364 sources within our sample of 421 targeted Galactic molecular clouds have been detected in both the C$^{18}$O and C$^{17}$O J=2-1 lines. All the detected IRAM and SMT spectra are presented in Figures \ref{fig:iram_both} and \ref{fig:smt_both}, respectively. In addition, the spectra of those sources without effective C$^{18}$O/C$^{17}$O determinations are shown in the \nameref{sec:app}.

\replaced{Assuming that the regions giving rise to C$^{18}$O and C$^{17}$O are the same, the abundance ratio $Ratio_{\rm corr}$ can be determined from the integrated intensity ratio I(C$^{18}$O)/I(C$^{17}$O) times the factor $(\nu\rm_{C^{17}O}/\nu\rm_{C^{18}O})^{2} = 1.047$, which comes from the frequency difference of both lines.}{{The abundance ratio $Ratio_{\rm corr}$ can be determined from the integrated intensity ratio I(C$^{18}$O)/I(C$^{17}$O) times the factor $(\nu\rm_{C^{17}O}/\nu\rm_{C^{18}O})^{2} = 1.047$ (see description in \citetalias{Zhang2020}).}} Our spectral fitting results, including the peak value ($T\rm_{peak}$), the integrated intensities of C$^{18}$O and C$^{17}$O with their uncertainties, and the abundance ratios with their uncertainty, are also presented in Table \ref{tab:results}.

\newpage
\setlength{\tabcolsep}{1.5mm}{
\startlongtable
\begin{deluxetable*}{cccccccccccc}
\tabletypesize{\footnotesize}
\tablecaption{Observational parameters and $^{18}$O/$^{17}$O isotope ratios.\label{tab:results}}
\tablehead{
\colhead{Source Name} & \colhead{R.A.} & \colhead{Decl.} & \colhead{$R_{GC}$} & \colhead{$d$} & \colhead{Telescope} & \colhead{Line} & \colhead{rms} & \colhead{$V_{peak}$} &
\colhead{$\int{T{\rm _{mb}d\nu}}$} & \colhead{$T\rm_{peak}$} & \colhead{$Ratio_{\rm corr}$}\\
\colhead{} & \colhead{(J2000)} & \colhead{(J2000)} & \colhead{(kpc)} & \colhead{(kpc)} & \colhead{} & \colhead{} & \colhead{(mK)} & \colhead{(km s$^{-1}$)}  & \colhead{(K km s$^{-1}$)} & \colhead{(K)} & \colhead{}\\
\colhead{(1)} & \colhead{(2)} & \colhead{(3)} & \colhead{(4)} & \colhead{(5)} &
\colhead{(6)} & \colhead{(7)} & \colhead{(8)} &
\colhead{(9)} & \colhead{(10)} & \colhead{(11)} & \colhead{(12)}
}
\startdata
WB89 312 & 00 02 41.3 & 64 34 04.01 & 11.39 & 4.60   & SMT   & C$^{18}$O  & 43    & -47.52 (0.02) & 1.25 (0.04) & 0.99  & 2.80 (0.21) \\
      &       &       &       &       & SMT   & C$^{17}$O  & 23    & -47.65 (0.06) & 0.47 (0.03) & 0.22  &  \\
WB89 325 & 00 14 26.6 & 64 28 30.29 & 10.57 & 3.39  & SMT   & C$^{18}$O  & 43    & -35.83 (0.01) & 5.36 (0.07) & 2.55  & 4.02 (0.12) \\
      &       &       &       &       & SMT   & C$^{17}$O  & 29    & -36.02 (0.03) & 1.40 (0.04) & 0.63  &  \\
WB89 326 & 00 15 29.1 & 61 14 40.99 & 10.76 & 3.69  & SMT   & C$^{18}$O  & 43    & -38.99 (0.01) & 2.97 (0.04) & 2.45  & 2.96 (0.10) \\
      &       &       &       &       & SMT   & C$^{17}$O  & 26    & -39.19 (0.03) & 1.05 (0.03) & 0.56  &  \\
WB89 330 & 00 20 58.1 & 62 40 18.01 & 10.78 & 3.66  & IRAM  & C$^{18}$O  & 59    & -38.99 (0.01) & 2.52 (0.05) & 2.59  & 3.83 (0.43) \\
      &       &       &       &       & IRAM  & C$^{17}$O  & 67    & -39.00 (0.04) & 0.69 (0.08) & 0.60  &  \\
WB89 331 & 00 21 19.4 & 63 19 19.99 & 11.81 & 5.02  & IRAM  & C$^{18}$O  & 252   & -51.69 (0.01) & 1.98 (0.05) & 2.30  & 4.21 (0.46) \\
      &       &       &       &       & IRAM  & C$^{17}$O  & 79    & -51.67 (0.04) & 0.49 (0.05) & 0.48  &  \\
WB89 336 & 00 26 55.8 & 65 10 27.80 & 14.07 & 7.72  & SMT   & C$^{18}$O  & 39    & -72.38 (0.03) & 0.58 (0.03) & 0.49  & 5.43 (0.97) \\
      &       &       &       &       & SMT   & C$^{17}$O  & 18    & -72.48 (0.08) & 0.11 (0.02) & 0.10  &  \\
WB89 344 & 00 33 42.0 & 66 49 49.01 & 14.25 & 7.87  & SMT   & C$^{18}$O  & 25    &       &       &       &  \\
      &       &       &       &       & SMT   & C$^{17}$O  & 37    &       &       &       &  \\
G121.29+00.65 & 00 36 47.3 & 63 29 02.18 & 8.86  & 0.93  & IRAM  & C$^{18}$O  & 35    & -17.50 (0.01) & 14.16 (0.06) & 5.17  & 3.24 (0.05) \\
      &       &       &       &       & IRAM  & C$^{17}$O  & 43    & -17.74 (0.02) & 4.58 (0.06) & 1.43  &  \\
  &   &   &    &    & SMT   & C$^{18}$O  & 36    & -17.48 (0.01) & 10.45 (0.07) & 3.95  & 3.48 (0.05) \\
      &       &       &       &       & SMT   & C$^{17}$O  & 25    & -17.66 (0.02) & 3.14 (0.04) & 1.00  &  \\
\enddata
\tablecomments{Column (1): source name. Columns (2) and (3): equatorial J2000 coordinates. Column (4): the galactocentric distance $R_{GC}$. Column (5): the heliocentric distance $d$. Column (6): telescopes used. Column (7): molecular species. Column (8): the rms (velocity resolutions 0.26 and 0.33 km s$^{-1}$ for IRAM and SMT, respectively) value in units of $T\rm_{mb}$. Column (9): \replaced{the velocity of each component.}{{the velocity of the peak $T\rm_{mb}$.}} Column (10): the integrated line intensity of C$^{18}$O and C$^{17}$O with standard deviation errors in parentheses. Column (11): the line peak values in $T\rm_{mb}$. Column (12): the frequency-corrected abundance ratio with its error in parentheses.\\
(This table is available in its entirety in machine-readable form.)}
\end{deluxetable*}
}

\section{Analysis and Discussion}\label{sec:dsc}
\subsection{Observational Effects}\label{sec:obseff}
In \citetalias{Zhang2020}, we have analyzed a series of factors that may affect the ratios derived from our J=1-0 data of C$^{18}$O and C$^{17}$O, including observational effects and chemical and physical factors. As already mentioned in \replaced{Section \ref{sec:intro}}{{\citetalias{Zhang2020}}}, chemical fractionation can be safely neglected due to the high first ionization potential of oxygen. However, optical depth effects have to be analyzed, since the J=2-1 line of C$^{18}$O normally shows a larger optical depth than its J=1-0 counterpart in a wide range of physical conditions \citep{Wouterloot2008}. We will check the degree of saturation in the C$^{18}$O J=2-1 line and assess its possible effect on our ratio results in Section \ref{sec:cp_1-0_2-1}. Here we investigate other potential biases related to our J = 2-1 data, following the analysis of the J = 1-0 data in \citetalias{Zhang2020}.

\begin{figure}[h]
\centering
\includegraphics[width=\textwidth]{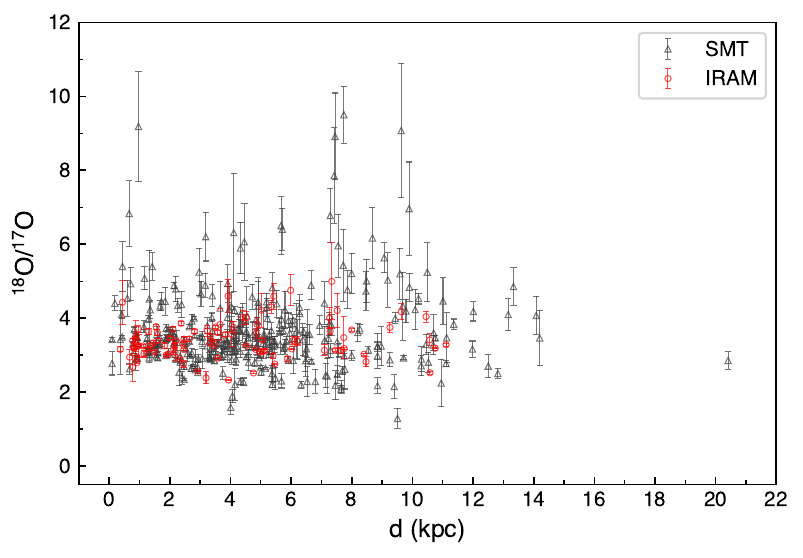}
\caption{Abundance ratios of the entire sample plotted against heliocentric distance. Red circles and black triangles represent our IRAM 30 m and SMT 10 m measurements, respectively.}
\label{fig:ratio_dsun}
\end{figure}

In Figure \ref{fig:ratio_dsun}, the isotopic ratios are plotted against heliocentric distance to investigate effects related to distance, i.e., to biases in beam dilution. No systematic dependence can be found, as in our J=1-0 data. This implies that the linear resolution of our data is not playing an important role, suggesting negligible small scale variations within the individual sources.

\begin{figure}[h]
\centering
\includegraphics[width=\textwidth]{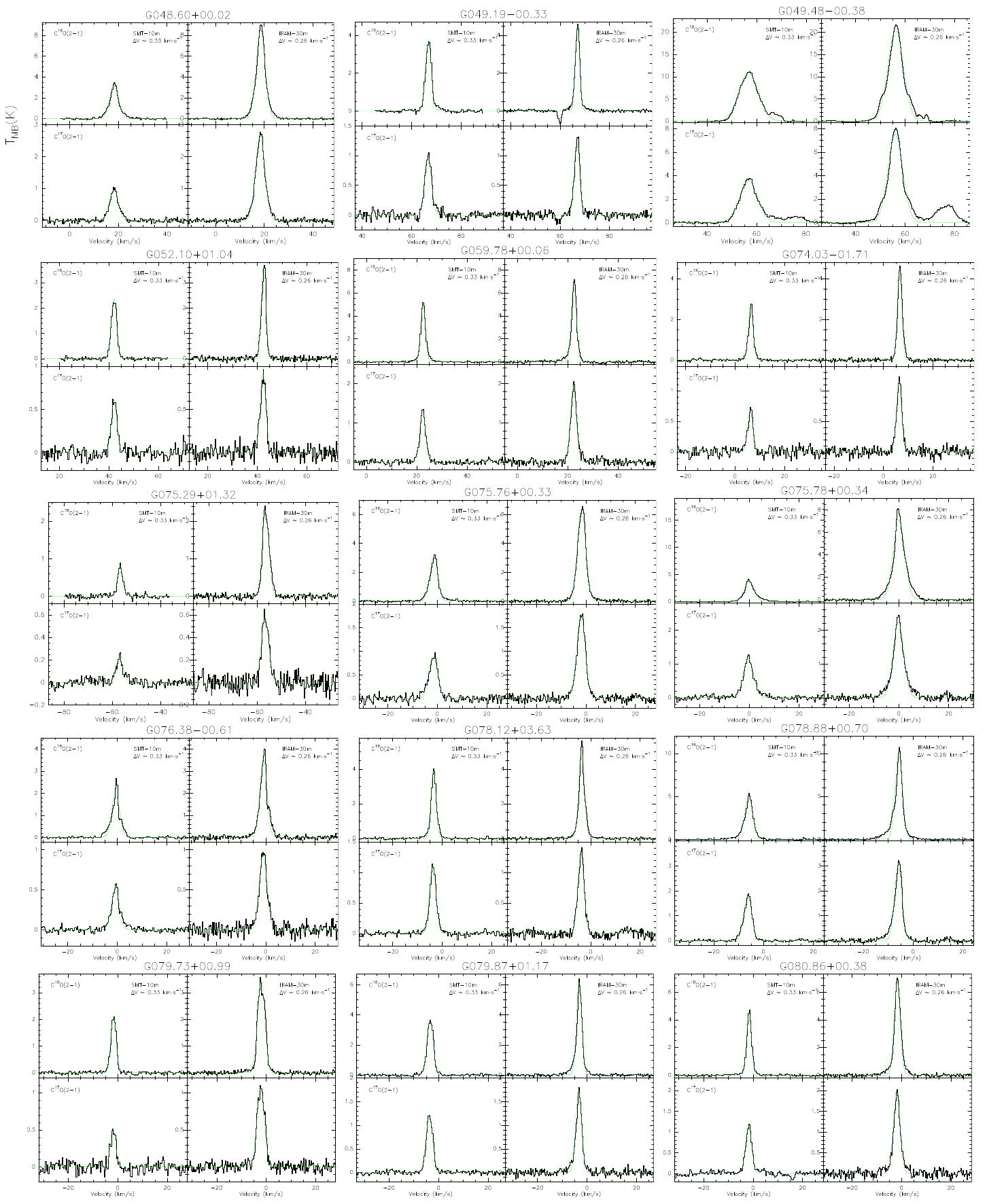}
\caption{The spectra of C$^{18}$O (upper panels) and C$^{17}$O (lower panels) with green fit lines of those 41 sources detected by both the SMT 10 m (left columns) and the IRAM 30 m (right columns) telescopes.\\
(An extended version of this figure is available).}
\label{fig:iram_smt}
\end{figure}

\begin{figure}[h]
\centering
\subfigure[]{ \label{fig:cp_size} %% label for first subfigure 
\includegraphics[width=8.5cm]{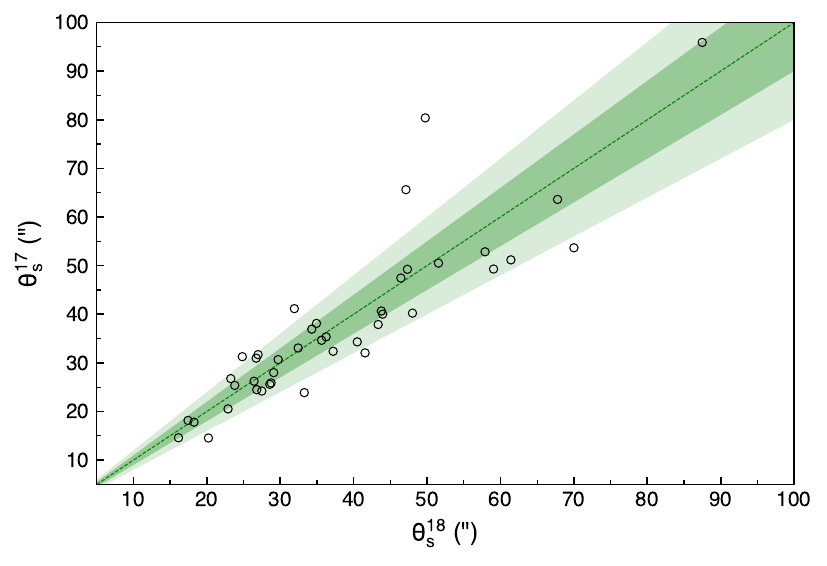}}
\hspace{0.5cm}
\subfigure[]{ \label{fig:cp_ratio} %% label for second subfigure 
\includegraphics[width=8.3cm]{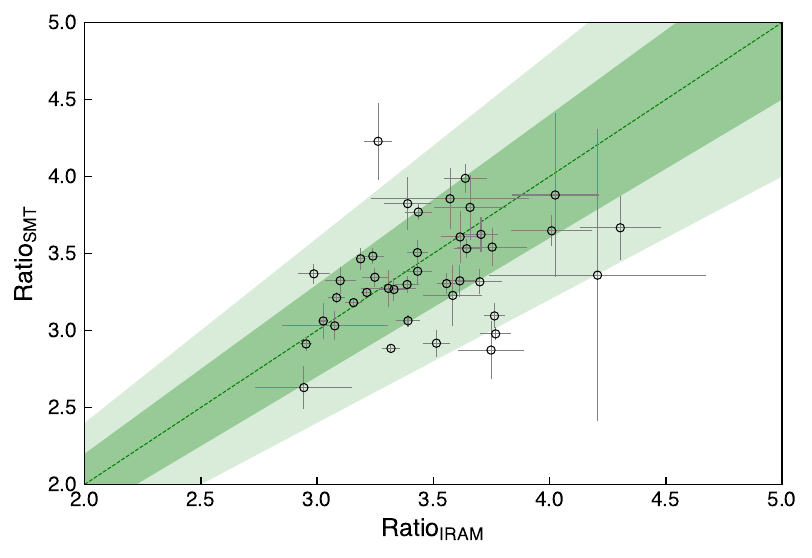}}
\caption{The comparison of estimated source sizes from C$^{18}$O and C$^{17}$O J=2-1 lines (left panel) and the isotope ratio measured by IRAM 30 m and SMT 10 m (right panel) for those 41 sources detected by both telescopes toward identical positions. The green dashed lines indicate $\theta_{s}^{18}$ = $\theta_{s}^{17}$ and Ratio${\rm _{IRAM}}$ = Ratio${\rm _{SMT}}$, while the dark and light green shaded areas indicate the $\pm$ 10\% and $\pm$ 20\% error ranges, respectively.} \label{fig:cp_size&ratio} %% label for entire figure
\end{figure}

A related analysis has been carried out making use of those sources measured by both the IRAM 30 m and the SMT 10 m telescopes, involving beam sizes of $\sim$10$\arcsec$ and $\sim$29$\arcsec$, which correspond to a ratio in covered areas by almost an order of magnitude. Here we also have to note that C$^{18}$O and C$^{17}$O were measured simultaneously by the IRAM 30 m (thanks to its broad bandwidth of $\sim$8 GHz) but were observed separately by the SMT 10 m (due to its narrow bandwidth of $\sim$0.25 GHz). Different observing conditions may lead to uncertainties. In case the lines are measured separately, pointing errors may affect peak and integrated intensities in different ways \citep{Wouterloot2008}. As mentioned in Section \ref{sec:d&r}, 57 sources out of 62 targets are detected by both the IRAM 30 m and the SMT 10 m, which could be used to quantify these effects. \replaced{However, because of 16 sources with targeted position deviations between both telescopes, only the remaining 41 sources are used for later analysis.}{{The coordinates of the observed positions were taken from \cite{Reid2014, Reid2019} and \cite{Wouterloot1989}. However 16 sources had targeted position differences between both telescopes due to errors in the input coordinates for one night observation, thus only the remaining 41 sources are used for later analysis.}} C$^{18}$O and C$^{17}$O J=2-1 spectra from these 41 sources are presented in Figure \ref{fig:iram_smt} and the spectral fit parameters are listed in Table \ref{tab:comp_SMT_IRAM}. All fitted lines detected with a smaller beam show larger main beam brightness temperatures ($T\rm_{mb}$) than those from the SMT 10 m with a larger beam. This indicates that the sizes of our sources are smaller than the SMT 10 m beam size and that beam dilution is not negligible. In this case, the intensity of detected lines should be corrected for the beam dilution effect and the brightness temperatures ($T_{B}$) of the sources can be derived from the main beam brightness temperature dilution:
\begin{equation}
T_{B}=T\rm_{mb}\frac{\theta_{s}^{2}+\theta_{beam}^{2}}{\theta_{s}^{2}},
\label{e:beamdilution}
\end{equation}
where $\theta_{s}$ and $\theta_{beam}$ are source size and beam size, respectively. With Equation \eqref{e:beamdilution}, we can estimate source size $\theta_{s}$ through beam sizes $\theta_{beam}$ of the two antennas and two measured $T\rm_{mb}$ values.
Both C$^{18}$O and C$^{17}$O line data are employed to estimate the size of our sample. We plotted their measured source sizes from C$^{17}$O lines against those from C$^{18}$O lines in Figure \ref{fig:cp_size}. It is found that the estimated sizes from both lines are mostly (\textgreater 80\%) comparable, within an error range of 20\% (see green shaded regions). Figure \ref{fig:cp_ratio} compares measured $^{18}$O/$^{17}$O ratios from both telescopes. It shows that the measured ratios by both telescopes are mostly consistent ($\sim$90\% sources, 37 out of 41), within a 20\% error range. To summarize, any observational bias is not significant with respect to our resulting $^{18}$O/$^{17}$O isotope ratios.

\setlength{\tabcolsep}{1.5mm}{
\startlongtable
\begin{deluxetable*}{cccccccccccc}
\tabletypesize{\footnotesize}
\tablecaption{A comparison of IRAM 30 m and SMT 10 m results for the 41 common sources with C$^{18}$O and C$^{17}$O detections.
\label{tab:comp_SMT_IRAM}}
\tablehead{
\colhead{Source Name} & \colhead{R.A.} & \colhead{Decl.} & \colhead{$R_{GC}$} & \colhead{$d$} & \colhead{Telescope} & \colhead{Line} & \colhead{rms} & \colhead{$V_{peak}$} &
\colhead{$\int{T{\rm _{mb}d\nu}}$} & \colhead{$T\rm_{peak}$} & \colhead{$Ratio_{\rm corr}$}\\
\colhead{} & \colhead{(J2000)} & \colhead{(J2000)} & \colhead{(kpc)} & \colhead{(kpc)} & \colhead{} & \colhead{} & \colhead{(mK)} & \colhead{(km s$^{-1}$)}  & \colhead{(K km s$^{-1}$)} & \colhead{(K)} & \colhead{}\\
\colhead{(1)} & \colhead{(2)} & \colhead{(3)} & \colhead{(4)} & \colhead{(5)} &
\colhead{(6)} & \colhead{(7)} & \colhead{(8)} &
\colhead{(9)} & \colhead{(10)} & \colhead{(11)} & \colhead{(12)}
}
\startdata
G121.29+00.65 & 00 36 47.3 & 63 29 02.18 & 8.86  & 0.93  & IRAM  & C$^{18}$O  & 35    & -17.50 (0.01) & 14.16 (0.06) & 5.17  & 3.24 (0.05) \\
      &       &       &       &       & IRAM  & C$^{17}$O  & 43    & -17.74 (0.02) & 4.58 (0.06) & 1.43  &  \\
      &       &       &       &       & SMT   & C$^{18}$O  & 36    & -17.48 (0.01) & 10.45 (0.07) & 3.95  & 3.48 (0.05) \\
      &       &       &       &       & SMT   & C$^{17}$O  & 25    & -17.66 (0.02) & 3.14 (0.04) & 1.00  &  \\
G122.01-07.08 & 00 44 58.4 & 55 46 47.60 & 9.67  & 2.17  & IRAM  & C$^{18}$O  & 32    & -50.21 (0.00) & 12.65 (0.05) & 4.80  & 3.08 (0.23) \\
      &       &       &       &       & IRAM  & C$^{17}$O  & 34    & -51.37 (0.07) & 4.30 (0.33) & 1.47  &  \\
      &       &       &       &       & SMT   & C$^{18}$O  & 44    & -50.83 (0.01) & 10.02 (0.05) & 4.15  & 3.03 (0.09) \\
      &       &       &       &       & SMT   & C$^{17}$O  & 76    & -51.03 (0.04) & 3.46 (0.11) & 1.25  &  \\
G123.06-06.30a & 00 52 24.7 & 56 33 50.51 & 10.15  & 2.82  & IRAM  & C$^{18}$O  & 43    & -30.30 (0.01) & 14.28 (0.08) & 3.86  & 3.64 (0.06) \\
      &       &       &       &       & IRAM  & C$^{17}$O  & 42    & -30.58 (0.03) & 4.10 (0.06) & 1.08  &  \\
      &       &       &       &       & SMT   & C$^{18}$O  & 29    & -30.44 (0.01) & 10.33 (0.05) & 2.82  & 3.53 (0.06) \\
      &       &       &       &       & SMT   & C$^{17}$O  & 26    & -30.78 (0.03) & 3.06 (0.05) & 0.76  &  \\
WB89 382 & 01 08 49.5 & 62 33 14.00 & 11.31  & 4.04  & IRAM  & C$^{18}$O  & 78    & -43.61 (0.01) & 3.33 (0.07) & 2.47  & 3.57 (0.34) \\
      &       &       &       &       & IRAM  & C$^{17}$O  & 80    & -43.74 (0.10) & 0.98 (0.09) & 0.40  &  \\
      &       &       &       &       & SMT   & C$^{18}$O  & 41    & -43.78 (0.01) & 3.01 (0.04) & 1.92  & 3.86 (0.20) \\
      &       &       &       &       & SMT   & C$^{17}$O  & 33    & -43.85 (0.05) & 0.82 (0.04) & 0.36  &  \\
G133.94+01.06 & 02 27 03.8 & 61 52 25.21 & 9.80  & 1.95  & IRAM  & C$^{18}$O  & 36    & -47.24 (0.02) & 32.75 (0.06) & 6.58  & 3.03 (0.02) \\
      &       &       &       &       & IRAM  & C$^{17}$O  & 37    & -47.57 (0.01) & 11.33 (0.06) & 2.14  &  \\
      &       &       &       &       & SMT   & C$^{18}$O  & 69    & -47.32 (0.01) & 28.39 (0.12) & 6.01  & 3.06 (0.11) \\
      &       &       &       &       & SMT   & C$^{17}$O  & 63    & -47.87 (0.09) & 9.71 (0.36) & 1.98  &  \\
\enddata
\tablecomments{Column (1): source name. Columns (2) and (3): equatorial
J2000 coordinates. Column (4): the galactocentric distance $R_{GC}$. Column (5): the heliocentric distance $d$. Column (6): telescopes used. Column (7): molecular species. Column (8): the rms (velocity resolutions are the same as in Table \ref{tab:results}) value in units of $T\rm_{mb}$. Column (9): \replaced{the velocity of each component.}{{the velocity of the peak $T\rm_{mb}$.}} Column (10): the integrated intensities of C$^{18}$O and C$^{17}$O with standard deviation errors in parentheses. Column (11): the line peak values in $T\rm_{mb}$. Column (12): the frequency-corrected abundance ratio with its error in parentheses.\\
(This table is available in its entirety in machine-readable form.)}
\end{deluxetable*}
}

\subsection{Galactic Interstellar $^{18}$O/$^{17}$O Gradient}\label{sec:gradient}
\subsubsection{A prominent $^{18}$O/$^{17}$O gradient from the J=2-1 lines of C$^{18}$O and C$^{17}$O}\label{sec:Pgradient}
\begin{figure}[h]
\centering
\subfigure[]{ \label{fig:ratio_dgc} %% label for first subfigure 
\includegraphics[width=8.5cm]{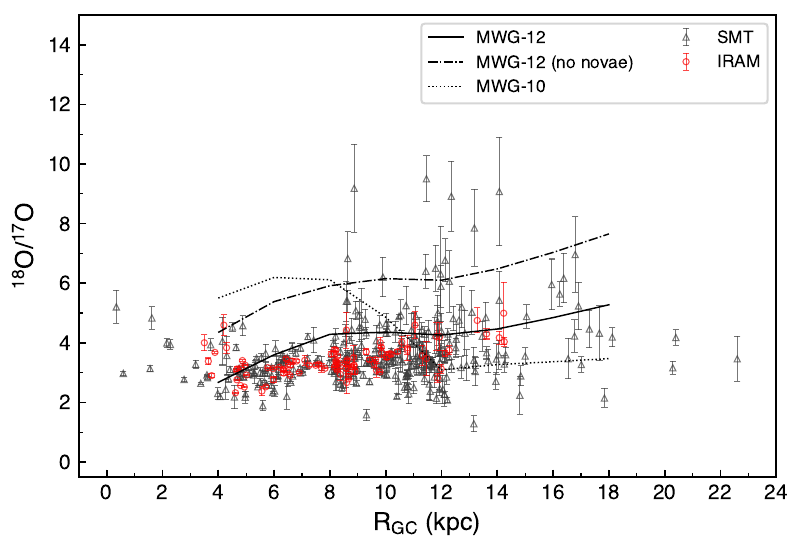}}
\hspace{0.5cm}
\subfigure[]{ \label{fig:ratio_dgc_bin} %% label for second subfigure 
\includegraphics[width=8.4cm]{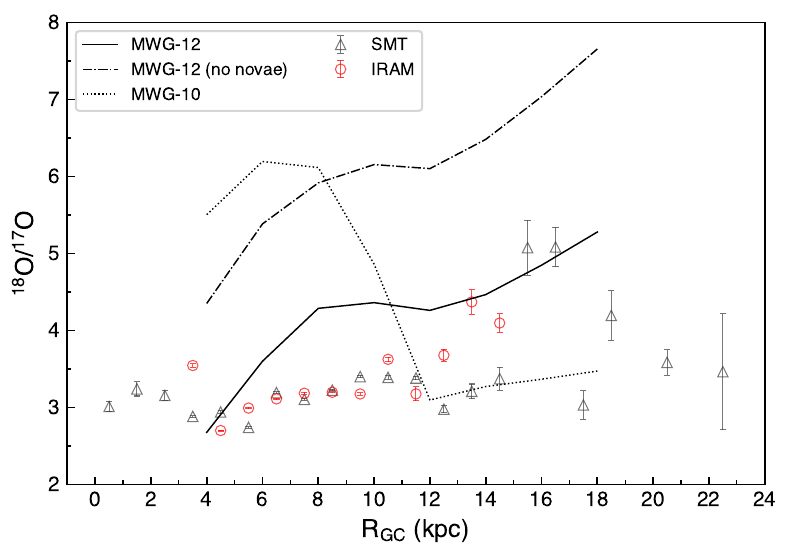}}
\caption{$^{18}$O/$^{17}$O isotope ratios plotted as a function of galactocentric distance, $R_{GC}$ (Figure \ref{fig:ratio_dgc}, left panel). In Figure \ref{fig:ratio_dgc_bin} (right panel), we plot the weighted average values (weight 1/$\sigma^2$, see \citetalias{Zhang2020}) of the ratio in bins of 1 kpc in $R_{GC}$. The red circles and black triangles are the results from our IRAM 30 m and SMT 10 m measurements, respectively. The curves represent predictions of the most recent galactic chemical evolution model; the dotted curve represents the model adopting the new yields by \cite{Limongi2018} for non-rotating stars (MWG-10), the dash-dotted and solid curves are for rotating stars without or with novae, respectively (MWG-12; see details in \citealt{Romano2019}).}\label{fig:ratio}
\end{figure}

With the IRAM 30 m and the SMT 10 m measurements of C$^{18}$O and C$^{17}$O J=2-1, we obtained $^{18}$O/$^{17}$O isotope ratios from 364 sources. Ratios are plotted against their galactocentric distances in Figure \ref{fig:ratio_dgc}. As in the case of our J = 1-0 results (see \citetalias{Zhang2020}), we find rising $^{18}$O/$^{17}$O ratios with increasing galactocentric distance, though there is significant scatter, especially for the SMT 10 m measurements. In order to show the trend more clearly, we average our results in bins of 1 kpc in galactocentric distance for both the SMT 10 m and the IRAM 30 m sample. The bin-averaged ratios are plotted as a function of galactocentric distance in Figure \ref{fig:ratio_dgc_bin}. Apparently, both data sets show similar gradients, i.e., the $^{18}$O/$^{17}$O ratios are smaller near the GC and larger in the outskirts of the Galaxy.

\begin{figure}[ht]
\centering
\includegraphics[width=\textwidth]{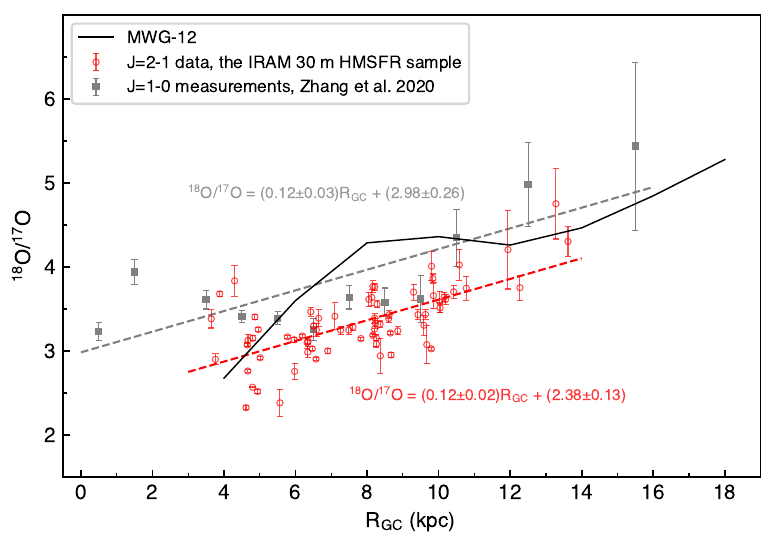}
\caption{A prominent radial $^{18}$O/$^{17}$O gradient is confirmed by our HMSFR sample, encompassing 72 sources with accurate distance values. Red and grey colors are used to indicate the results from J=2-1 and J=1-0 data (the latter taken from \citetalias{Zhang2020}), respectively. The dashed lines represent unweighted linear fits, while the expressions give the fitting parameters. As in Figure \ref{fig:ratio_dgc}, the curve represents predictions of the GCE model including both rotating stars and novae (MWG-12 in \citealt{Romano2019}).}
\label{fig:iram_zhang}
\end{figure}

\begin{figure}[ht]
\centering
\includegraphics[width=\textwidth]{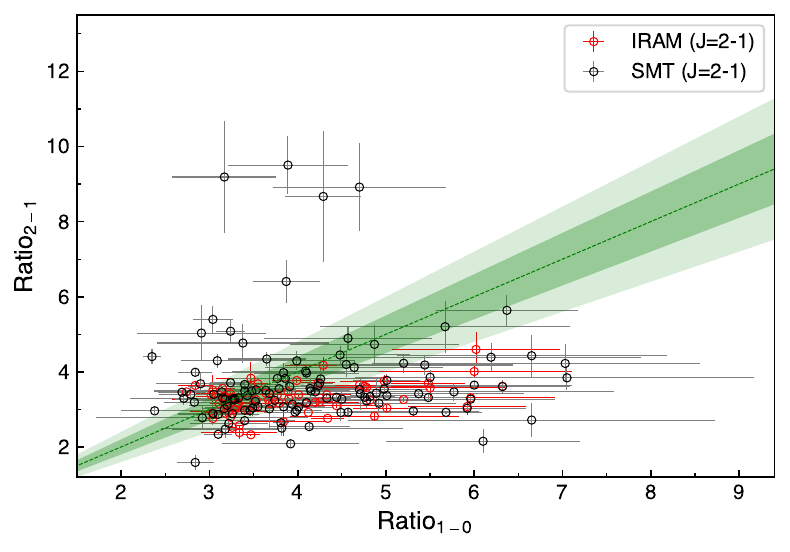}
\caption{A comparison of isotope ratios from our J=1-0 (\citetalias{Zhang2020}) and J=2-1 (this paper) data.\added{{ Red and grey colors indicate IRAM and SMT J=2-1 data, respectively.}} For the green line and shaded areas, see the caption to Figure \ref{fig:cp_size&ratio}.}
\label{fig:cp_1-0_2-1a}
\end{figure}

Only accounting for the HMSFR subsample with accurate distance values, measured by trigonometric parallaxes making use of maser lines \citep{Reid2014, Reid2019}, statistical results should be particularly reliable. In addition, IRAM 30 m measurements should be more accurate, thanks to the wide bandwidth (both C$^{18}$O and C$^{17}$O line were observed simultaneously) and smaller beam size (less beam dilution, leading to higher line temperatures and signal-to-noise ratios), which minimizes observational effects (see details in Section \ref{sec:obseff}). Thus we took the HMSFR sample with parallax distances and IRAM 30 m measurements to determine accurately the $^{18}$O/$^{17}$O gradient in terms of the galactocentric distance. This subsample contains 72 sources, covering a galactocentric distance range of 3-14 kpc. Their $^{18}$O/$^{17}$O ratios as a function of galactocentric distance are plotted in Figure \ref{fig:iram_zhang}. In contrast to the large scatter for the entire sample (Figure \ref{fig:ratio_dgc}), Figure \ref{fig:iram_zhang} shows the correlation of $^{18}$O/$^{17}$O with galactocentric distance more clearly. The unweighted linear fit provides a radial gradient of the ratio, $^{18}$O/$^{17}$O = (0.12 $\pm$ 0.02)$R_{GC}$ + (2.38 $\pm$ 0.13), with a Pearson's rank correlation coefficient of $R=0.67$. We also present the bin-averaged results from our J=1-0 measurements \citepalias{Zhang2020} in Figure \ref{fig:iram_zhang}. Our results from both J=1-0 and J=2-1 data show identically one significant correlation between the ratio and the galactocentric distance, with the same slope for both fit lines (Figure \ref{fig:iram_zhang}). In addition we find that ratios from the J=2-1 lines tend to be systematically smaller than those from the J=1-0 data. For those 142 sources with measured ratios from J=1-0 and J=2-1 data, the ratio from J=1-0 data (Ratio$_{1-0}$) is mostly larger than that from J=2-1 data (Ratio$_{2-1}$; see Figure \ref{fig:cp_1-0_2-1a}) in $\sim$80\% of the sources. The difference in the measured ratios from the two transitions will be discussed in the subsection \ref{sec:cp_1-0_2-1}.

Modeling works on galactic chemical evolution (GCE) can certainly help us to better understand the galactic radial gradient of $^{18}$O/$^{17}$O. Similar to \citetalias{Zhang2020}, modeling results from the latest theoretical works \citep{Romano2019} are shown in Figure \ref{fig:ratio_dgc} and Figure \ref{fig:iram_zhang} (solid, dotted and dashed-dotted curves), including MWG-10 (massive stars without fast rotation), MWG-12 (including both rotating massive stars and novae) and MWG-12nn (including rotating massive stars, but no nova contribution). As shown in Figure \ref{fig:ratio_dgc} and Figure \ref{fig:iram_zhang}, MWG-12 can match best our measurements, i.e., including contributions from both fast rotating massive stars and novae. Contributions from fast rotators can provide high $^{18}$O/$^{17}$O ratios in the outer, metal-poor Galactic disk, which plays a unique role in the positive gradient of $^{18}$O/$^{17}$O (e.g., \citealt{Frischknecht2016}). And the contribution of nova nucleosynthesis should not be neglected, although it still involves many uncertainties \citep{Romano2017}. Large variations of the individual nova yields coupled to the rarity of nova outbursts may be partly responsible for the scatter presented in our data. However, the difference between the J=2-1 data and the modeling results is larger than the corresponding difference related to the J=1-0 results (see Figure \ref{fig:iram_zhang}). This may be caused by the larger optical depth of the C$^{18}$O J=2-1 lines, which will be discussed in the next subsection.

\subsubsection{Comparison of J=2-1 and J=1-0 measurements}\label{sec:cp_1-0_2-1}
\begin{figure}[h]
\centering
\includegraphics[width=\textwidth]{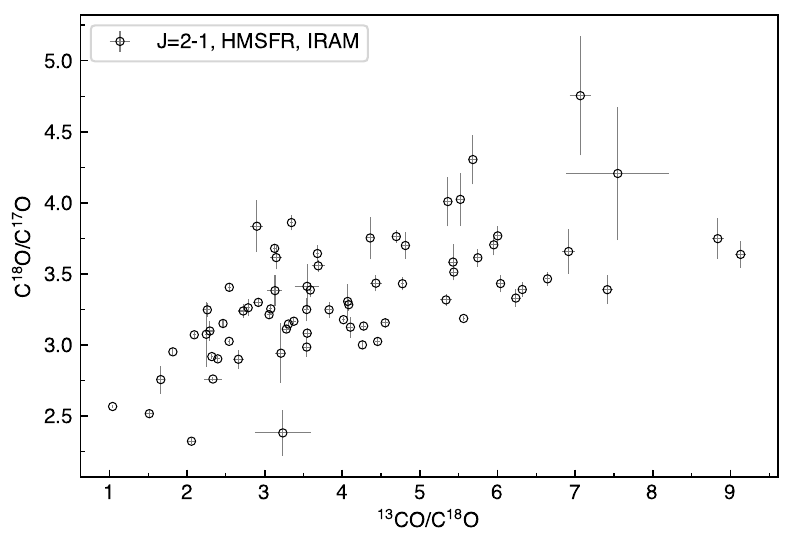}
\caption{The C$^{18}$O/C$^{17}$O integrated intensity ratio as a function of $^{13}$CO/C$^{18}$O for the J=2-1 transition.}
\label{fig:13co}
\end{figure}

In Section \ref{sec:Pgradient}, we found that the ratios from J=2-1 data tend to be systematically lower than those from the J=1-0 measurements and those from modeling (MWG-12, Figure \ref{fig:iram_zhang}).
This is probably related to the optical depths of the C$^{18}$O J=2-1 lines. The optical depth of the C$^{17}$O line can be directly determined by fitting its hyperfine structure lines (see details in \citetalias{Zhang2020}), and then we can get the optical depth of the C$^{18}$O line, assuming that it is about four times that of the corresponding C$^{17}$O line. However, the velocity differences between the hyperfine components of our C$^{17}$O J=2-1 lines ($\sim$1.2 km s$^{-1}$) is in most cases too small to be spectroscopically resolvable, thus leading to unreliable results. As substitution, a qualitative analysis to test for saturation effects of C$^{18}$O was made by making use of the more abundant isotopic species $^{13}$CO, whose J=2-1 line was also covered by our IRAM 30 m observations. Apparently, any source in which the intensity of the C$^{18}$O line is reduced by opacity effects can be expected to present a correspondingly greater diminution in its $^{13}$CO spectrum. Thus, we can expect the greatest saturation effects in the C$^{18}$O/C$^{17}$O ratios in those sources showing the lowest $^{13}$CO/C$^{18}$O line intensity ratios \citep{Penzias1981}. $^{13}$CO/C$^{18}$O against C$^{18}$O/C$^{17}$O is presented in Figure \ref{fig:13co} and we can find that small $^{13}$CO/C$^{18}$O ratios are associated with small C$^{18}$O/C$^{17}$O values. This implies that our measured C$^{18}$O J=2-1 lines are to some degree saturated.

\begin{figure}[ht]
\centering
\includegraphics[width=\textwidth]{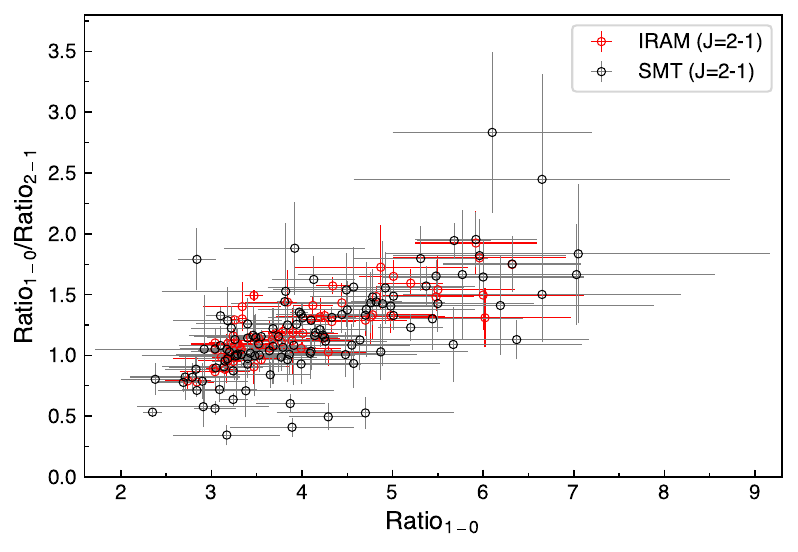}
\caption{The Ratio$_{1-0}$/Ratio$_{2-1}$ values are plotted as a function of Ratio$_{1-0}$.\added{{ Red and grey colors indicate IRAM and SMT J=2-1 data, respectively.}}}
\label{fig:cp_1-0_2-1b}
\end{figure}

For those 142 sources with both Ratio$_{1-0}$ and Ratio$_{2-1}$, we plotted Ratio$_{1-0}$/Ratio$_{2-1}$ as a function of Ratio$_{1-0}$ in Figure \ref{fig:cp_1-0_2-1b}. We can find that the Ratio$_{1-0}$/Ratio$_{2-1}$ slowly increases with Ratio$_{1-0}$, which means that in absolute but not in relative terms (e.g., Section \ref{sec:Pgradient} and Figure \ref{fig:iram_zhang}) Ratio$_{2-1}$ has a lower increasing rate than Ratio$_{1-0}$. This suggests a more pronounced saturation in the C$^{18}$O J=2-1 line with regard to the C$^{18}$O J=1-0 line and thus the real abundance ratio may be underestimated by our analysis of the J=2-1 lines. Therefore, it is necessary to check the degree of saturation in the C$^{18}$O J=2-1 line and assess its possible effect on the ratio results.

\begin{figure}[h]
\centering
\subfigure[]{ \label{fig:radexA} %% label for first subfigure 
\includegraphics[width=8.5cm]{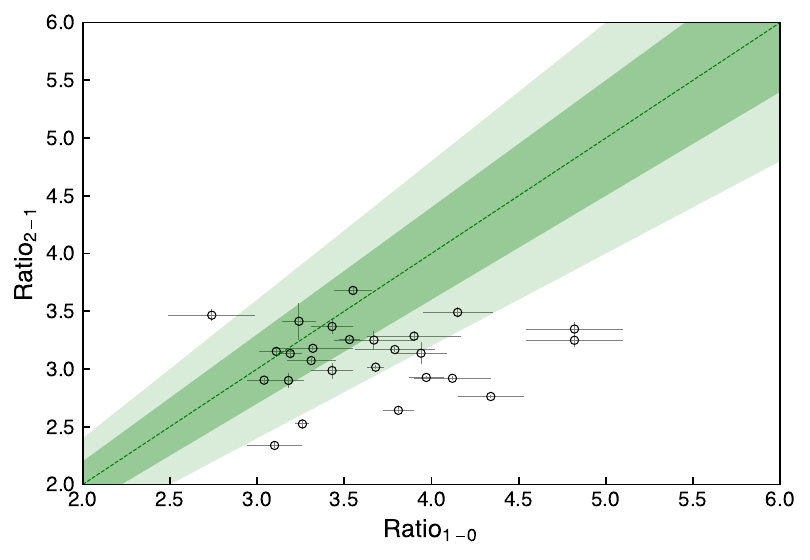}}
\hspace{0.5cm}
\subfigure[]{ \label{fig:radexB} %% label for second subfigure 
\includegraphics[width=8.5cm]{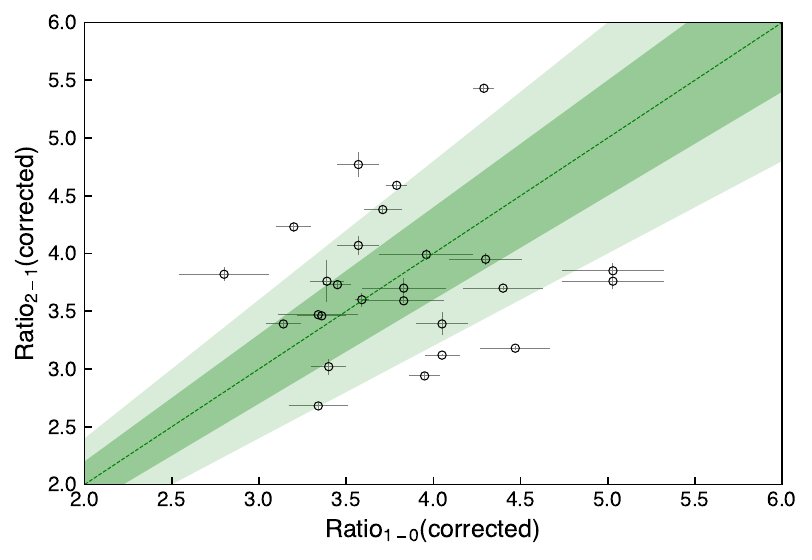}}
\caption{A comparisons of the $^{18}$O/$^{17}$O ratios from our J=1-0 and J=2-1 data for 24 sources. Left and right panel are before and after correction of RADEX calculations, respectively. The green line and shaded areas are similar as Figure \ref{fig:cp_size&ratio}.}
\label{fig:radex}
\end{figure}

In order to give a quantitative estimation of the saturation effect, we performed a non-LTE analysis for the C$^{18}$O J=2-1 line using the RADEX software \citep{Van2007} toward our sample. We cross-matched our sample with the published ATLASGAL catalogue \citep{Billington2019} and obtained $n(\rm H_2)$ data for 24 sources. The kinetic temperatures of these 24 sources were also collected \citep{Milam2005, Hill2010, Dunham2011, Urquhart2011, Wienen2012, Cyganowski2013, Svoboda2016, Billington2019, Chen2021, Wienen2021}. With $n(\rm H_2)$, $T_{k}$, molecular data from the Leiden Atomic and Molecular Database\footnote{https://home.strw.leidenuniv.nl/~moldata/} (LAMDA) and our spectral line data, we performed RADEX calculations and obtained the optical depths for the C$^{18}$O J=2-1 lines. It ranges from 0.1 to 2.0, mostly (\textgreater 80\%) less than 1.0, with an average value of $\sim$0.5 (corresponding correction of $\sim$20\% on the ratio). This indicates that in some cases C$^{18}$O J=2-1 is saturated, but overall the opacity effect should be moderate. For comparison, we also calculated the optical depths of the C$^{18}$O J=1-0 line for those 24 sources. With regard to the C$^{18}$O 2-1 line, smaller optical depths were obtained for the J=1-0 transition, with an average optical depth of $\sim$0.1 (corresponding correction of $\sim$4\%), which agrees with results in \citetalias{Zhang2020}, i.e., non-significant optical depth effects for the J=1-0 line. After correcting for the optical depth effect in both lines, we find that the ratio results measured by both J=2-1 and J=1-0 transitions are consistent within errors (see Figure \ref{fig:radexB}).

\section{Summary}\label{sec:sum}
We are performing systematic observations of multi-transition lines of C$^{18}$O and C$^{17}$O toward a large sample of molecular clouds in the Galaxy, covering the Galactic plane from the central region to the far outer Galaxy ($\sim$22 kpc), to search for variations in the oxygen $^{18}$O/$^{17}$O isotope ratio as a function of galactocentric distance. In this work, we present a large C$^{18}$O and C$^{17}$O J=2-1 survey, observed with the IRAM 30 m and the SMT 10 m telescopes. 96 out of 103 sources were successfully detected in both C$^{18}$O and C$^{17}$O J=2-1 lines through the IRAM 30 m, while 325 out of 380 sources were detected in both lines using the SMT 10 m. 62 sources were observed by both telescopes, 41 of them were detected toward exactly the same position in both C$^{18}$O and C$^{17}$O J=2-1, which is used to check the quality of the data and to evaluate beam dilution. Our main results are as follows:

(1) From the measured J=2-1 line intensities, $^{18}$O/$^{17}$O ratios could be obtained for 364 sources. No systematic variation is found between the isotopic ratio and the heliocentric distance, which implies that any observational bias due to beam dilution is not significant. We analyzed observational data from those sources detected by both the SMT 10 m and IRAM 30 m telescopes, derived source sizes and obtained consistent $^{18}$O/$^{17}$O isotope ratios. This also implies that linear resolution and beam dilution do not significantly affect the determined isotope ratios.

(2) The measured $^{18}$O/$^{17}$O isotope ratios of 364 sources tend to increase with galactocentric distance, though the scatter is large. The gradient is most clearly seen in the subsample of high-mass star-forming regions (HMSFRs) detected with the IRAM 30 m telescope. These data are characterized by high signal-to-noise ratios and are also the least affected by observational effects. The unweighted linear fit determines the radial gradient as $^{18}$O/$^{17}$O = (0.12 $\pm$ 0.02)$R_{GC}$ + (2.38 $\pm$ 0.13), with a Pearson's rank correlation coefficient of $R=0.67$. This is consistent with prediction by a recent model of Galactic chemical evolution, considering contributions from both fast rotating massive stars and novae. The randomness of the latter's yield may cause a part of the scatter revealed by our data.

(3) Our measured $^{18}$O/$^{17}$O ratios from the C$^{18}$O and C$^{17}$O J=2-1 lines (Ratio$_{2-1}$) tend to be lower than those from the J=1-0 lines (Ratio$_{1-0}$), though both present identical trends of increasing $^{18}$O/$^{17}$O ratios with rising galactocentric distance. The difference is likely caused by the fact that in many instances the C$^{18}$O J=2-1 line is not optically thin. This is supported by the positive correlation between Ratio$_{1-0}$/Ratio$_{2-1}$ and Ratio$_{1-0}$, and the RADEX non-LTE model calculation results, which suggest average optical depths of about 0.5 and 0.1 for C$^{18}$O J=2-1 and 1-0, respectively. After optical depth correction, consistent $^{18}$O/$^{17}$O ratio can be obtained within the limits of observational accuracy.

\begin{acknowledgements}
\section*{ACKNOWLEDGMENTS}
This work is supported by the National Key R\&D program of China (2022YFA1603102) and the Natural Science Foundation of China (No. 12373021, 12041302, 11590782). D.R. thanks the Italian National Institute for Astrophysics (INAF) for the support provided to the project \textit{“An in-depth theoretical study of CNO element evolution in galaxies”} via Theory Grant 2022, Fu.~Ob.~1.05.12.06.08. We thank the operators and staff at both IRAM 30 m and SMT 10 m stations for their assistance during our observations.
\end{acknowledgements}

\section*{Appendix}
\label{sec:app}
The spectra of the targets that were not detected (or with only hints of a detection, which are candidates for additional observation) are shown in Figure \ref{fig:other}.

\setcounter{table}{0}
\setcounter{figure}{0}
\renewcommand{\thetable}{A\arabic{table}}
\renewcommand*{\theHtable}{\thetable}
\renewcommand{\thefigure}{A\arabic{figure}}
\renewcommand*{\theHfigure}{\thefigure}

\begin{figure*}[ht]
\centering
\includegraphics[width=\textwidth,page=1]{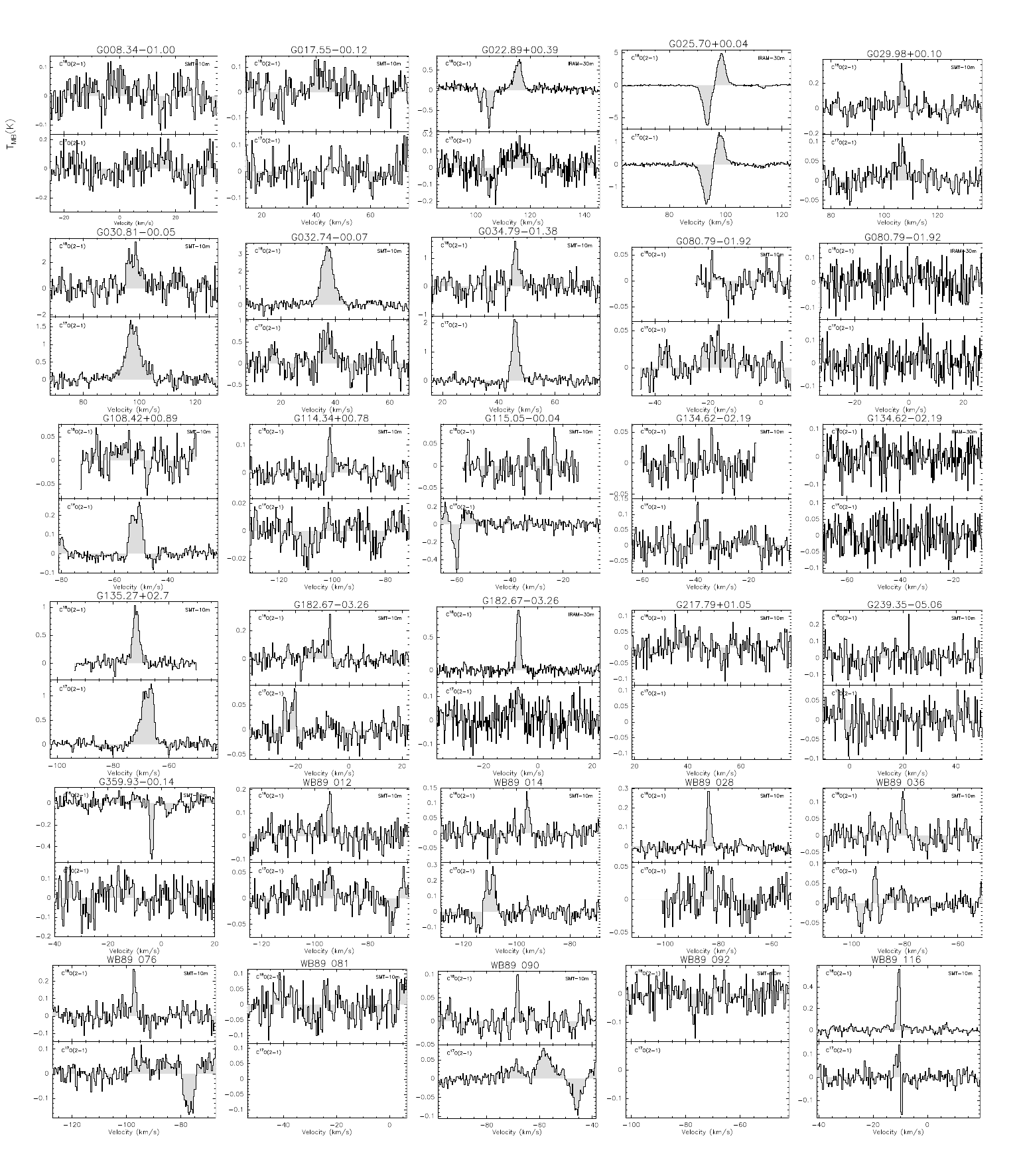}
\caption{Spectra of those sources without effective measurements of C$^{18}$O/C$^{17}$O.{\\
(An extended version of this figure is available).}}
\label{fig:other}
\end{figure*}

\clearpage
\bibliography{zou2021}{}
\bibliographystyle{aasjounal}
\end{CJK*}
\end{document}